\documentclass[aps,showpacs,nofootinbib,superscriptaddress]{revtex4-1}
\usepackage{epsf}
\usepackage{graphicx}
\usepackage{amsmath}
\def\tstrut{\vrule height2.5ex depth0pt width0pt} % used in tables
\def\slashchar#1{\setbox0=\hbox{$#1$}
   \dimen0=\wd0 \setbox1=\hbox{/} \dimen1=\wd1
   \ifdim\dimen0>\dimen1 \rlap{\hbox to \dimen0{\hfil/\hfil}} #1
   \else  \rlap{\hbox to \dimen1{\hfil$#1$\hfil}} / \fi}

\begin{document}

\title{Coherent pion production
off nuclei at T2K and MiniBooNE energies revisited} 

%\author{J.~E.~Amaro}
%\affiliation{Departamento de F{\'\i}sica At\'omica, Molecular y
%Nuclear,\\ Universidad de Granada, E-18071 Granada, Spain} 
\author{E. Hern\'andez} 
\affiliation{Departamento
de F\'\i sica Fundamental e IUFFyM,\\ Universidad de Salamanca, E-37008
Salamanca, Spain.}  
\author{J.~Nieves} 
\affiliation{ Instituto de F\'\i sica Corpuscular (IFIC), Centro Mixto
  CSIC-Universidad de Valencia, Institutos de Investigaci\'on de
  Paterna, Aptd. 22085, E-46071 Valencia, Spain} 
\author{M.~Valverde} 
\affiliation{Research Center for Nuclear Physics (RCNP), Osaka
University, Ibaraki 567-0047, Japan}


\today

\begin{abstract}
 As a result of a new improved fit to old bubble chamber
 data of the dominant axial
 $C_5^A$ nucleon-to-Delta form factor, and due to the relevance of
 this form factor for neutrino induced coherent pion production, we
 re-evaluate our model predictions in Phys.\ Rev.\ D {\bf 79}, 013002
 (2009) for different observables of the latter reaction. Central
 values for the total cross sections increase by 20\%$\sim$30\%, while
 differential cross sections do not change their shape
 appreciably. Furthermore, we also compute the uncertainties on total,
 differential and flux averaged cross sections induced by the errors
 in the determination of $C_5^A$.  Our new
 results turn out to be compatible within about $1\sigma$ with the
 former ones. Finally, we
 stress the existing tension between the recent experimental
 determination of the $\frac{\sigma({\rm CC coh}\pi^+)}{\sigma({\rm NC
 coh}\pi^0)}$ ratio by the SciBooNE Collaboration and the theoretical
 predictions.
 \end{abstract}

\maketitle

\section{Introduction}

Experimental analyses of neutrino induced coherent pion production  generally rely on the
 Rein--Sehgal (RS) model~\cite{rein,rein2} which is based on the partial
 conservation of the axial current (PCAC) hypothesis. In
 the RS model the pion-nucleus coherent cross section is written in
 terms of the pion-nucleon elastic cross section by means of
 approximations that are valid for high neutrino energies and small
 values of the nucleus momentum transfer
 square  and  of the lepton momentum transfer square ($q^2$).  As pointed out in
 Refs.~\cite{Amaro:2008hd,Hernandez:2009vm}, those approximations are
 less reliable for neutrino energies below/around 1 GeV, light nuclei,
 like carbon or oxygen, and finite values of $q^2$. These are the
 energies and targets used in present and forthcoming neutrino
 oscillation experiments.

There are other approaches to coherent production that do not rely on
PCAC but on microscopic models for pion production at the nucleon
level~\cite{Amaro:2008hd,Fogli:1979cz,
kelkar,Sato:2003rq,singh,luis1,luis2,Nakamura:2009iq,Nakamura:2009kr,Martini:2009uj}. 
The dominant contribution to
the elementary amplitude at low energies is given by the $\Delta$-pole
mechanism ($\Delta$ excitation and its subsequent decay into $\pi
N$). Medium effects on the $\Delta$ mass and
width, final pion distortion, as well as
nonlocalities in the pion momentum, are very important and are taken
into account in microscopic calculations.  Similarly to PCAC models,
the process is dominated by the axial part of the weak current and it
is thus very sensitive to nucleon-to-Delta axial form factors.

Very recently, the role of nonlocalities in the $\Delta$ momentum has
also been investigated~\cite{Leitner:2009ph,Praet:2009zz,Nakamura:2009iq,Nakamura:2009kr}. In
Ref.~\cite{Leitner:2009ph} it is claimed that their neglect, the so called, local
approximation, leads to an overestimate of the coherent production
cross section that can be as large as a factor of 2 for neutrino
energies of 500\,MeV.  Similar results were obtained in
Ref.~\cite{Praet:2009zz}. Final
pion distortion and  in medium modifications of the $\Delta$
properties were not considered, and it was not clear whether those approximations could not
affect the results. Final pion distortion and in medium modification
of the $\Delta$ properties were included in
Refs.~\cite{Nakamura:2009iq,Nakamura:2009kr}, were nonlocal effects on the $\Delta$
momentum were incorporated in the $\Delta$ self-energy in the
first-order approximation. They also observe a large reduction in the
total cross section due to the nonlocal aspects of the $\Delta$
propagation in the medium. However, as claimed by the authors of
Ref.~\cite{Nakamura:2009iq}, this that not mean that earlier
microscopic calculations
~\cite{Amaro:2008hd,Fogli:1979cz,
kelkar,Sato:2003rq,singh,luis1,luis2,Martini:2009uj} are wrong, as there $\Delta$
nonlocal effects are taken into account in an effective way through
the in medium modification of the $\Delta$ properties which were
fitted to observables.  
 
In the model we developed in Ref.~\cite{Amaro:2008hd}, 
the $\Delta$ was treated in the local
approximation. However, the modifications of the $\Delta$ in medium
properties are such that similar models give a good reproduction of
pionic atoms and $\pi-$nucleus 
scattering~\cite{GarciaRecio:1989xa,Nieves:1993ev}, pion
photoproduction~\cite{Carrasco:1989vq}, pion
electroproduction~\cite{Gil:1997bm}, ($^3$He,t)
~\cite{FernandezdeCordoba:1992df} and
elastic $\alpha-$proton~\cite{FernandezdeCordoba:1993az}
reactions.
We share the claim of the authors of
Ref.~\cite{Nakamura:2009iq} and we believe this treatment of the
$\Delta$, where certainly non-local effects are being effectively
(partially) taken into account, is also adequate for neutrino induced reactions. Nevertheless,
this interesting issue deserves future investigations.

Our model in Ref.~\cite{Amaro:2008hd} is based on a microscopic model
at the nucleon level, described in detail in
Ref.~\cite{Hernandez:2007qq}, that, besides the dominant
$\Delta$ pole contribution, takes into account background terms required 
by chiral symmetry.
As a result of the inclusion of background terms, we had to re adjust the strength of the
dominant $\Delta$ pole contribution. The least known ingredients of
the model are the axial nucleon-to-$\Delta$ transition form
factors
of
which $C_5^A$, not only gives the largest contribution, but it also
controls all other axial form factors if one assumes Adler's
model~\cite{Adler:1968tw} that gives 
$C_4^A(q^2) = -\frac{C_5^A(q^2)}{4}, \
C_3^A(q^2) = 0 $, 
and  PCAC is used to obtain  $C_6^A(q^2) = C_5^A(q^2)
\frac{M^2}{m_\pi^2-q^2}$.
This strongly suggested to us the readjustment of $C_5^A$ to the
experimental data.

Information on pion production off the nucleon comes mainly from 
two bubble chamber
experiments, ANL~\cite{anlviejo,anl,anl-cam73} and BNL~\cite{bnlviejo,bnl}. 
Assuming, as proposed in Ref.~\cite{Pa04}, the $q^2$ dependence
$
C_5^A(q^2) = \frac{C_5^A(0)}{(1-q^2/M^2_{A\Delta})^2}\,(
1-\frac{q^2}{3M_{A\Delta}^2})^{-1}
$,
we fitted in Ref.~\cite{Hernandez:2007qq} the flux-averaged $\nu_\mu
p\to \mu^- p \pi^+$ ANL $q^2$-differential cross section
for pion-nucleon invariant masses $W < 1.4$
GeV~\cite{anl,anlviejo} obtaining $C_5^A(0) = 0.867 \pm 0.075$ and 
$M_{A\Delta}=0.985\pm 0.082\,{\rm}$,
with a Gaussian correlation coefficient $r=-0.85$ and a
$\chi^2/dof=0.4$. The fitted axial mass was in good agreement with the
estimates of about 0.95 GeV and 0.84 GeV given in Refs.~\cite{anl,Pa05}.
 On the other
hand, the $C_5^A(0)$ value is some 30\% smaller than the prediction
obtained from the off diagonal Goldberger-Treiman relation (GTR) that
gives $\left.C_5^A(0)\right|_{GTR}=1.2$.  $C_5^A(0)$ is not
constrained by chiral perturbation theory ($\chi$PT) and lattice
calculations are still not conclusive about the size of possible
violations of the GTR. For instance, though values for $C_5^A(0)$ as
low as 0.9 can be inferred in the chiral limit from the results of
Ref.~\cite{negele07-prd}, they also predict
$C_5^A(0)/\left.C_5^A(0)\right|_{GTR}$ to be
greater than one.

Recently, two re-analysis have been carried out trying to make
compatible the GTR prediction for $C_5^A(0)$ and ANL data. In
Ref.~\cite{Leitner:2008ue}, $C_5^A(0)$ is kept to its GTR value and
three additional parameters, that control the $C_5^A(q^2)$ fall off,
are fitted to the ANL data.  Although ANL data are well reproduced, we find the
outcome in ~\cite{Leitner:2008ue} to be unphysical, as it
provides a quite pronounced $q^2-$dependence  giving rise to a too
large axial transition radius
of around 1.4 fm (further details 
 are discussed in \cite{Hernandez:2009zg}).

A second re-analysis~\cite{Graczyk:2009qm} brings in the discussion
two interesting points. First that both ANL and BNL data were measured
in deuterium, and second, the uncertainties in the neutrino flux
normalization. It is claimed in Ref.~\cite{Graczyk:2009qm}, that the latter
 could be responsible for 
 BNL total cross sections being systematically
larger than ANL ones.  In  Ref.~\cite{Graczyk:2009qm}
the authors do a combined best fit to the ANL and BNL data including
deuteron effects, which they evaluate as in Ref.~\cite{AlvarezRuso:1998hi}, and
 flux normalization uncertainties, treated as systematic
errors, and taken to be 20\% for ANL
data and 10\% for BNL data. With a pure dipole dependence for $C_5^A$, 
they found $C_5^A(0)=1.19 \pm 0.08$, in agreement with
the GTR estimate.

The works in Refs.~\cite{Leitner:2008ue,Graczyk:2009qm} consider only 
the $\Delta $ pole mechanism
but ignore the sizable non-resonant contributions
which are of special relevance for  neutrino energies below 1 GeV.  When
background terms are considered, the tension between ANL data and the
GTR prediction for $C_5^A(0)$ substantially increases as the results in
Ref.~\cite{Hernandez:2007qq} clearly shows. 

In our  work in Ref.~\cite{Hernandez:2010bx} we have performed a
fit to both ANL and BNL data in which: {\it i)} We have
included  the BNL total $\nu_\mu p \to \mu^- p \pi^+$ cross
section measurements of Ref.~\cite{bnlviejo}. We have just included the three lowest neutrino energies: 0.65, 0.9
and 1.1 GeV, since there is no
cut in the outgoing pion-nucleon invariant mass in the BNL data, and
we want to avoid heavier resonances from playing a significant role.  We have not used the BNL measurement of the
$q^2-$differential cross section, since it lacked an absolute
normalization.  {\it ii)} We have taken into account deuteron effects,
 {\it iii)}  the
uncertainties in the  neutrino flux normalizations,  20\% for ANL and 10\% 
for BNL data, are treated as fully
correlated systematic errors, improving thus the simpler treatment
adopted in Ref.~\cite{Graczyk:2009qm}, and finally {\it iv)} in some
fits, we have relaxed Adler's model constraints, in order to extract some direct
information on $C_{3,4}^A(0)$. For simplicity we took
$C_5^A(q^2) = \frac{C_5^A(0)}{(1-q^2/M^2_{A\Delta})^2}$.
As in Ref.~\cite{Graczyk:2009qm},
the consideration of BNL
data and flux uncertainties increased the value of $C_5^A(0)$ by about
9\%, while strongly reduced the statistical correlations between
$C_5^A(0)$ and $M_{A\Delta}$. The inclusion of background
terms reduced $C_5^A(0)$ by about 13\%, and deuteron effects increased
it by about 5\%, consistently with the results of
\cite{Hernandez:2007qq} and \cite{AlvarezRuso:1998hi,Graczyk:2009qm},
respectively. Fitted data was quite insensitive to
$C_{3,4}^A(0)$. 

In our most robust fit in Ref.~\cite{Hernandez:2010bx}  we used  
Adler's  constraints,  and we obtained
$C_5^A(0) = 1.00 \pm 0.11, \ M_{A\Delta}=0.93\pm 0.07\,{\rm
  GeV}$,
with a small Gaussian correlation coefficient $r=-0.06$ and a
$\chi^2/dof=0.32$. This violation of the  GTR is about
15\%, and it is smaller than that suggested in
Ref.~\cite{Hernandez:2007qq}, though it
is definitely greater than that claimed in Ref.~\cite{Graczyk:2009qm},
mostly because in Ref.~\cite{Graczyk:2009qm} background terms were not
considered. However, the GTR value and the $C_5^A(0)$  above
differ in less than $2\sigma$, and the discrepancy is even smaller if
Adler's constraints are removed.

 These new results are quite relevant for the neutrino induced
coherent pion production in nuclei which is a low $q^2$ dominated
reaction.  Background term
contributions to coherent production largely cancel for symmetric
nuclei~\cite{Amaro:2008hd} making the $\Delta$ pole mechanism the
unique contribution. As the process is dominated by the 
axial part of the weak current, it is 
 very sensitive to 
$C_5^A(0)$.  Thus, we would expect the results in
Ref.~\cite{Amaro:2008hd}, based in the determination of $C_5^A(0)$ of
Ref.~\cite{Hernandez:2007qq}, to underestimate cross sections by some
30\%. In this work we re-evaluate different pion coherent production
observables using our model of Ref.~\cite{Amaro:2008hd} but with the
new parameterization and results for $C_5^A$ obtained in 
Ref.~\cite{Hernandez:2010bx}. As the correlation coefficient is small in
this case, we shall treat the theoretical errors that derive from the
uncertainties in $C_5^A(0)$ and $M_{A\Delta}$ as independent and we
shall add them in quadratures.

\section{New results}\label{sec:results}
\begin{figure}[htb]
\begin{center}
\makebox[0pt]{\includegraphics[scale=0.239]{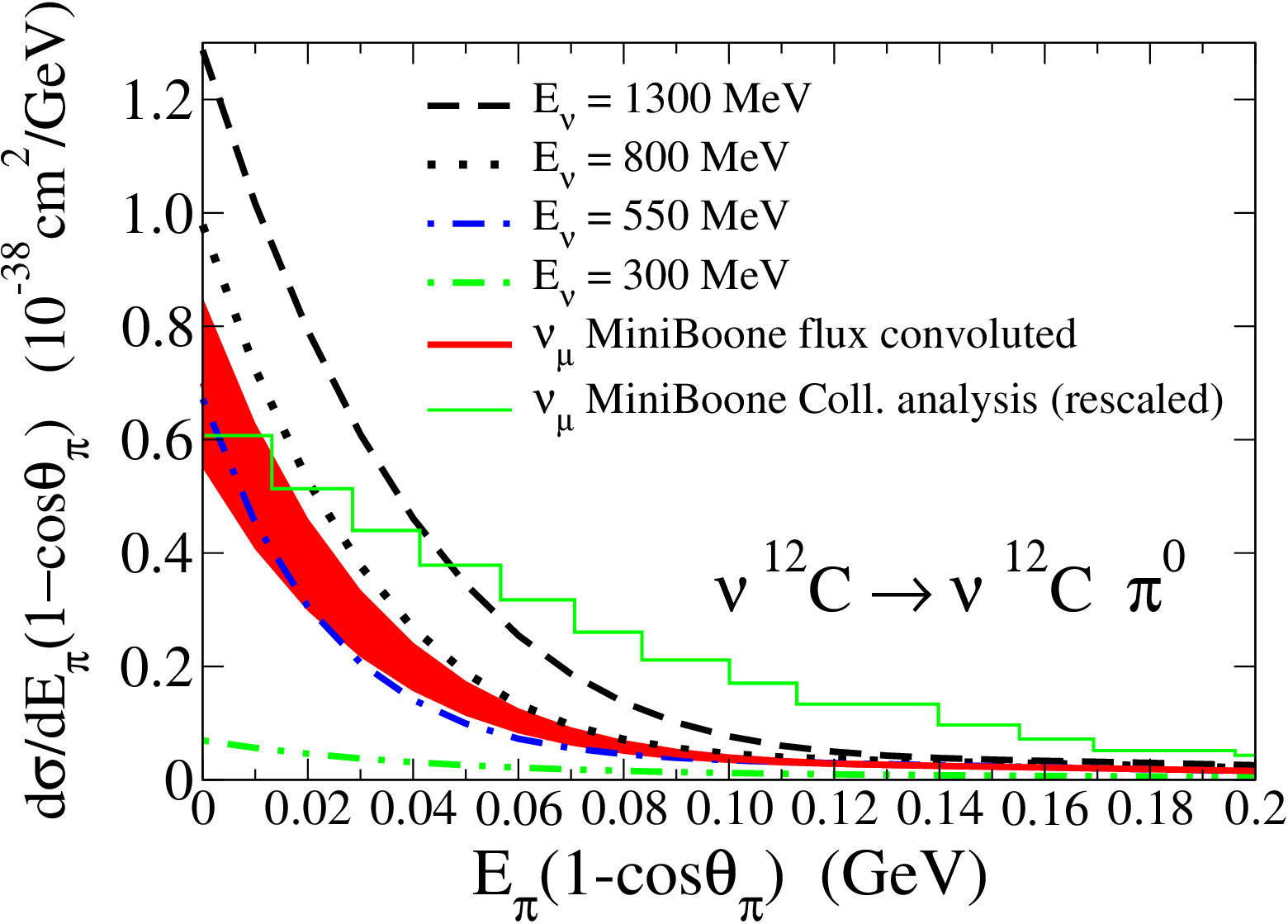}\hspace{.3cm}
\includegraphics[scale=0.239]{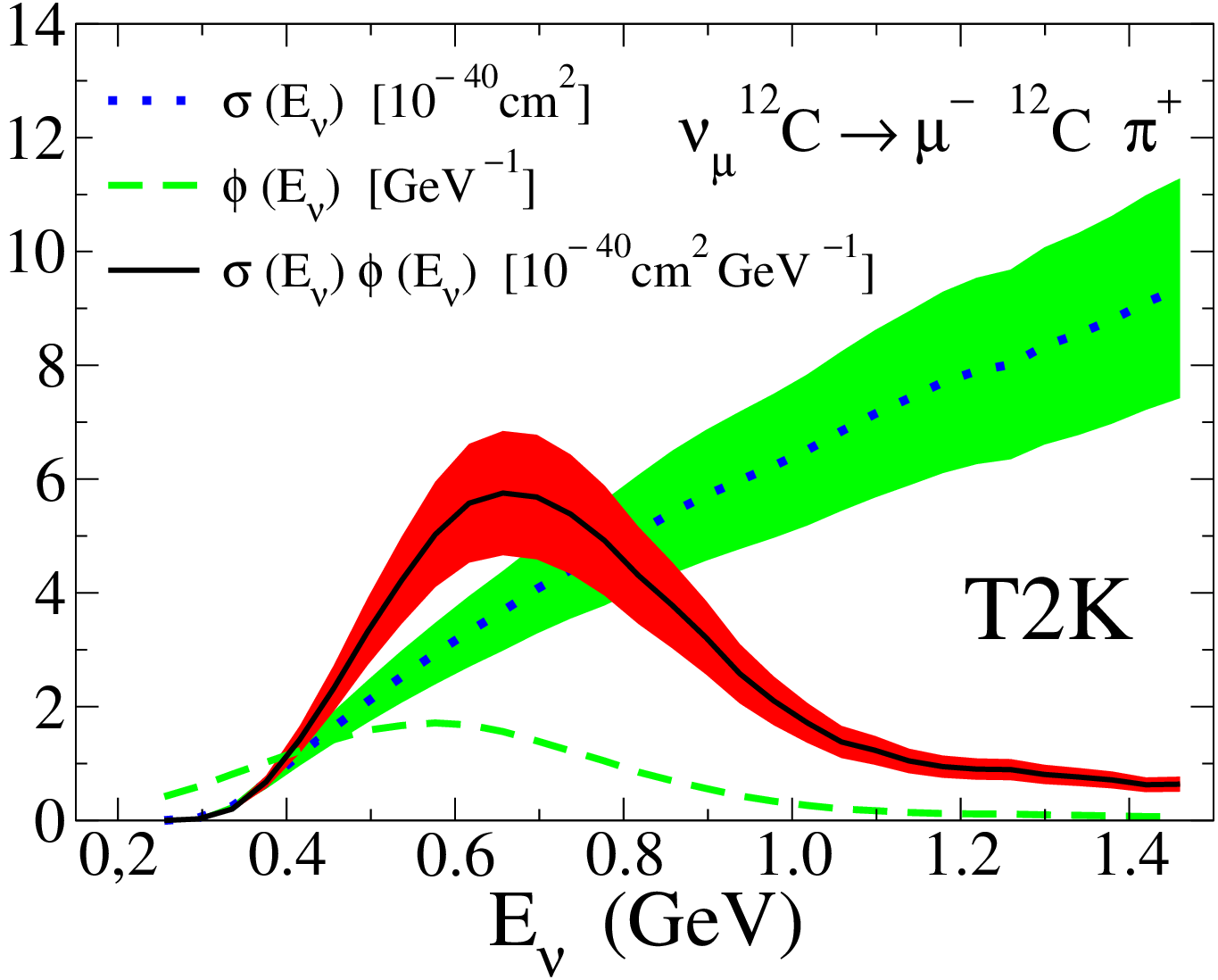}
\hspace{.375cm}
              \includegraphics[scale=0.239]{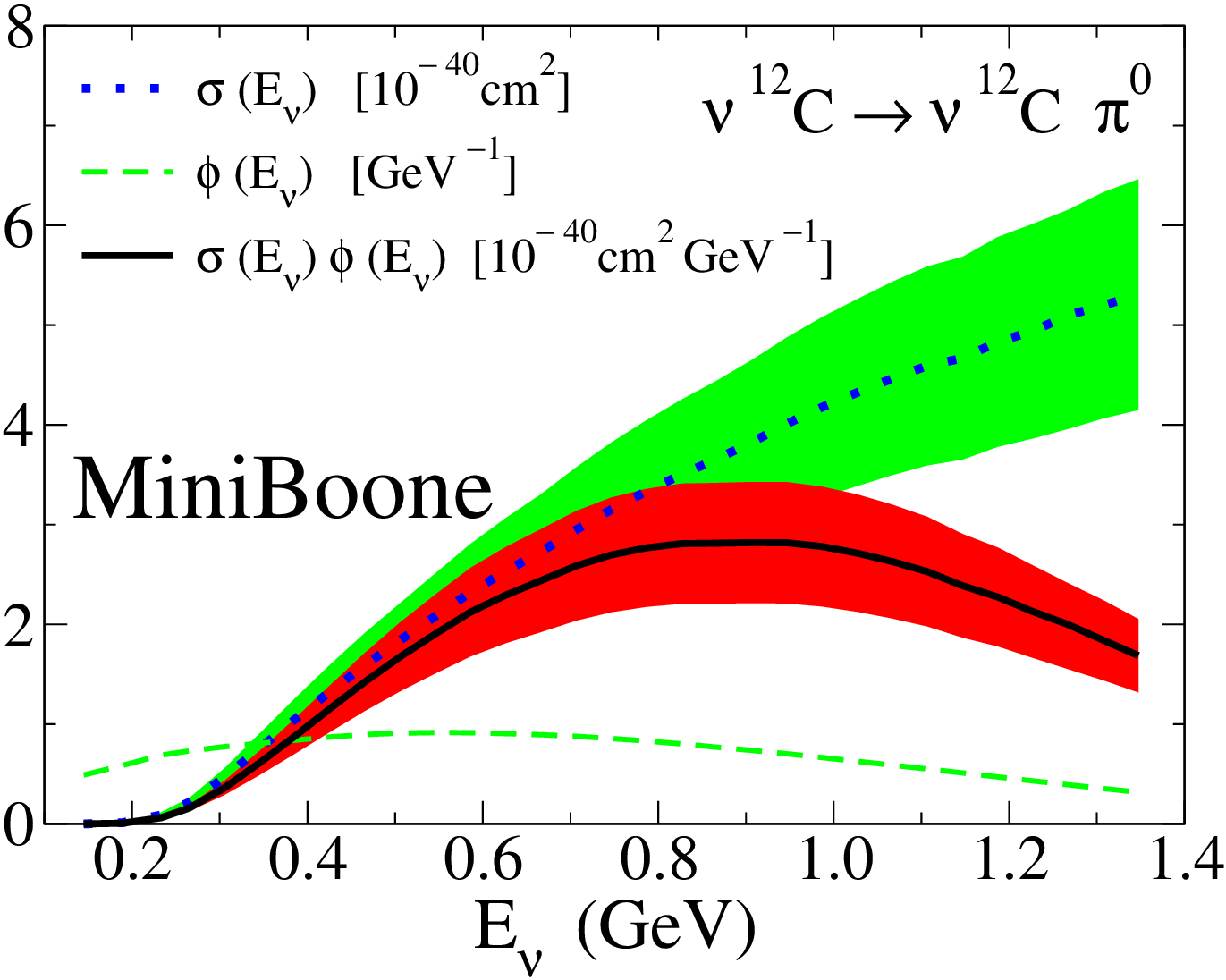}}\\
\end{center}
\caption{\footnotesize Left panel: Laboratory $E_\pi (1-\cos\theta_\pi)$
  distribution for the $\nu\, ^{12}{\rm C}\to \nu\, ^{12}{\rm C}\,
  \pi^0 $ reaction, at MiniBooNE energies. We also show, with the
  corresponding error band, the  distribution
  convoluted with the $\nu_\mu$ MiniBooNE flux.  We
  display the MiniBooNE published histogram, taken from the right
  panel of Fig.3 in Ref.~\protect\cite{AguilarArevalo:2008xs},
  conveniently scaled down so that their first bin matches our result
  at $E_\pi (1-\cos\theta_\pi)=0.005$~\,GeV. Middle and right panels: CC  and NC  coherent pion
 production cross sections in carbon (dots). We also show (solid
 lines) predictions multiplied by the T2K (middle) and MiniBooNE (right)
 $\nu_\mu$  energy spectra.  The dashed curves curves stand
 for the T2K and MiniBooNE $\nu_\mu$ fluxes normalized to one. Error
 bands are shown for our results.\\
 %\vspace{0.5cm}
 }\label{fig:flux-cross}
\end{figure}

We start by showing in the left panel of Fig.~\ref{fig:flux-cross} the differential cross
section with respect to the $E_\pi(1-\cos\theta_\pi)$ variable for
neutral current (NC) coherent $\pi^0$ production on carbon. $E_\pi$ is
the pion energy in the laboratory frame (LAB) while $\theta_\pi$ is
the LAB angle between the pion and the incoming neutrino. The shapes
of the distributions are completely similar to the ones we obtained in
Ref.~\cite{Amaro:2008hd}, but the absolute values increase by some
20\%$\sim$ 30\% depending on the neutrino energy. This is generally
true for other differential cross sections that we do not show
here. For the distribution convoluted with the MiniBooNE flux we find
an increase in the total cross section of about 29\%.
\begin{table}\begin{center}
    \begin{tabular}{lcccccc}\hline\tstrut
Reaction                 & Experiment &\hspace*{1cm}$\bar \sigma$\hspace*{1cm}&\ \ 
$\sigma_{\rm exp}$\ \ &
$E_{\rm max}^i $ ~~ & $\int_{E_{\rm low}^i}^{E_{\rm max}^i} dE \phi^i(E)
\sigma(E)$~~ & $\int_{E_{\rm low}^i}^{E_{\rm max}^i} dE \phi^i(E)$
\\
&  & [$10^{-40}$cm$^2$] & [$10^{-40}$cm$^2$] & [GeV]
&[$10^{-40}$cm$^2$]  \\\hline\tstrut
CC\phantom{*} $\nu_\mu + ^{12}$C    & K2K        &   $6.1\pm1.3$    &
$<7.7 $~\cite{Hasegawa:2005td}   &1.80&$5.0\pm1.0$&0.82         \\
CC\phantom{*} $\nu_\mu + ^{12}$C    & MiniBooNE  &   $3.8\pm0.8$    
&          &1.45&$3.5\pm0.7$&0.93 \\

CC\phantom{*} $\nu_\mu + ^{12}$C    & T2K        &   $3.2\pm0.6$   
 &          &1.45&$2.9\pm0.6$&0.91           \\
CC\phantom{*} $\nu_\mu + ^{16}$O    & T2K        &   $3.8\pm0.8$    &    &
  1.45& $3.4\pm0.7$&  0.91           \\
NC\phantom{*} $\nu_{{\mu}} + ^{12}$C    & MiniBooNE  & $2.6\pm0.5$ & $7.7\pm1.6\pm3.6$~\cite{Raaf}
&1.34&$2.2\pm0.5$&0.89\\
NC\phantom{*} $\nu_{{\mu}} + ^{12}$C    & T2K        & $2.3\pm0.5$    
 &     &1.34 &$2.1\pm0.5$
&0.90 \\
NC\phantom{*} $\nu_{{\mu}} + ^{16}$O    & T2K        & 
  $2.9\pm0.6$      & &  1.35     &  $2.6\pm0.6$ &  0.90  \\
CC\phantom{*} $\bar\nu_\mu + ^{12}$C    & T2K        &   $2.6\pm0.6$ &
&    1.45      & $1.8\pm0.4$ & 0.67          \\
NC\phantom{*} $\bar\nu_{{\mu}} + ^{12}$C    & T2K        &
 $2.0\pm0.4$     &     &1.34 &$1.3\pm0.3$
&0.64 \\
\hline
    \end{tabular}
  \end{center} 
  \caption{\footnotesize NC/CC $\nu_\mu$ and $\bar\nu_\mu$ coherent
    pion production total cross sections, with errors, for K2K,
    MiniBooNE and T2K experiments. In the case of CC K2K, the
    experimental threshold for the muon momentum $|\vec{k}_\mu|>$
    450 MeV is taken into account.    Details on the flux convolution are compiled in
    the last three columns. }
\label{tab:res} 
\end{table}

In Table~\ref{tab:res} we show our new predictions for, both NC 
and charged current (CC)
processes, for the
K2K~\cite{Hasegawa:2005td} and MiniBooNE~\cite{AguilarArevalo:2008xs}
flux averaged cross sections as well as for the future T2K
experiment.  In the middle and right panels of Fig.~\ref{fig:flux-cross}, we 
show some results for T2K
and MiniBooNE experiments. In all cases, the 
flux $\phi$ is normalized to one. As in Ref.~\cite{Amaro:2008hd}, and
since we neglect all resonances above the $\Delta(1232)$, we have set
up a maximum neutrino energy ($E_{\rm max}^i$) in the flux
convolution, approximating the convoluted cross section 
by 
$\bar \sigma \approx {\int_{E_{\rm low}^i}^{E_{\rm max}^i} dE \phi^i(E)
\sigma(E)}{ /}{\int_{E_{\rm low}^i}^{E_{\rm max}^i} dE \phi^i(E)} $,
where we fixed the upper limit in the integration  to $E_{\rm max}=1.45\,$GeV and 1.34 GeV for CC
and NC $\nu_\mu/\bar\nu_\mu$ driven processes, respectively.
$E_{\rm low}^i$ is the lower flux limit.
For the K2K case a threshold of 450 MeV for muon
momentum is also implemented~\cite{Hasegawa:2005td}  and  we 
can to go up to $E_{\rm max}^{\rm
CC,K2K}$=1.8 GeV. We  cover about 90\% of
the total flux in most of the cases. For the T2K antineutrino flux, we
cover just about 65\%, and therefore our results
 are less reliable.

Our central value cross sections increase by some 23\%$\sim$30\%, while the
errors associated to the uncertainties in the $C_5^A(0)$ and
$M_{A\Delta}$ determination are of the order of 21\%. Our new results
are thus compatible with former ones in Ref.~\cite{Amaro:2008hd}
within $1\sigma$. 
 Our prediction for the K2K experiment lies more
than 1$\sigma$ below the K2K upper bound, while we still predict an NC
MiniBooNE cross section notably smaller than that given in the PhD
thesis of J.L. Raaf~\cite{Raaf}. Note however, that  the
MiniBooNE Collaboration has not given an official value for the total
coherent cross section yet, and only the ratio
coherent/(coherent+incoherent) has been
presented~\cite{AguilarArevalo:2008xs}.  For the future T2K
experiment, we now get cross sections of the order
2.4$-$3.2$\times10^{-40}$cm$^2$ in carbon and about 2.9$-$3.8 $\times
10^{-40}$cm$^2$ in oxygen.

\begin{figure}[htb]
\begin{center}
\makebox[0pt]{\includegraphics[scale=0.25]{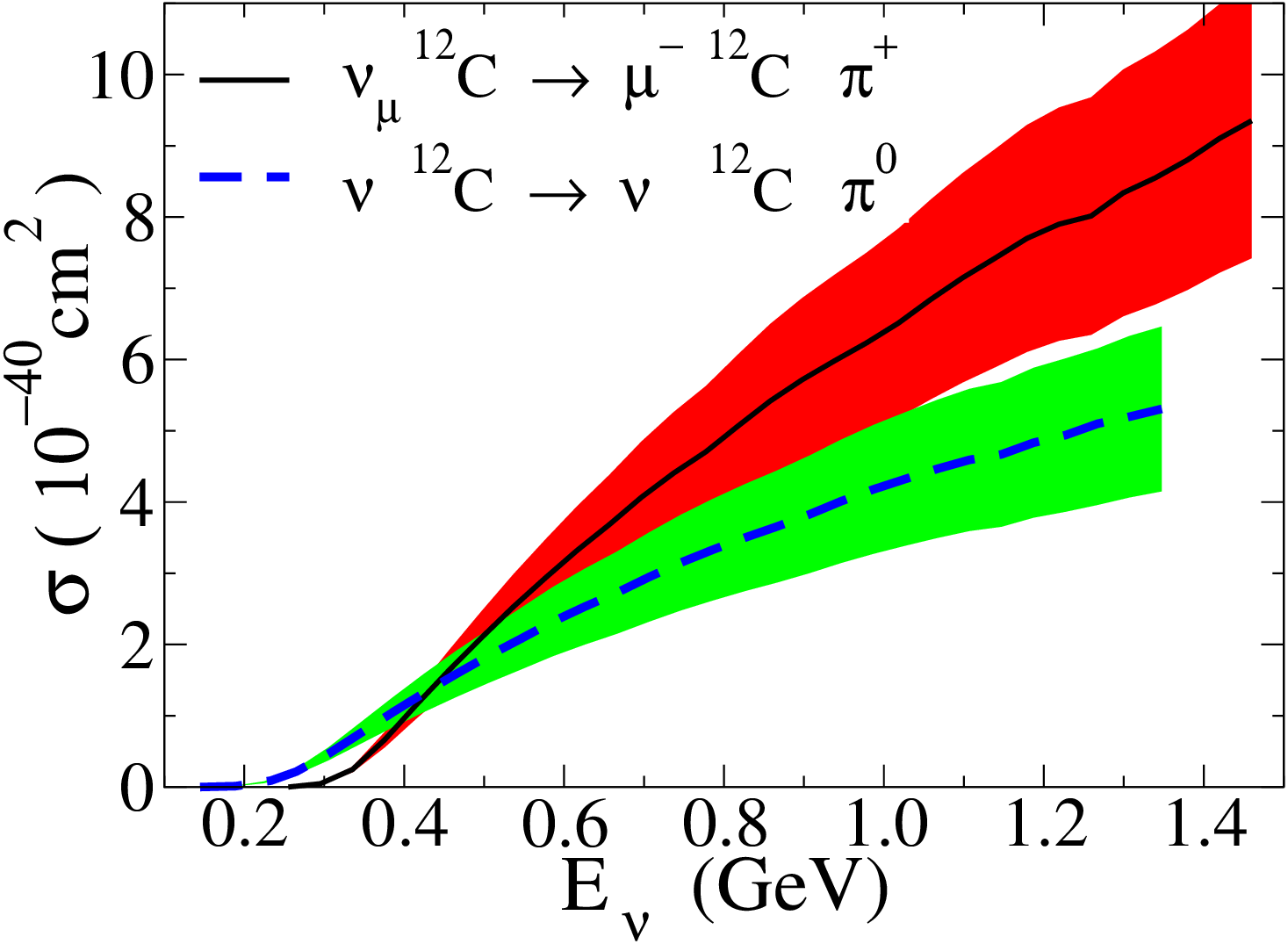}
\hspace{.1cm}
             
\includegraphics[scale=0.25]{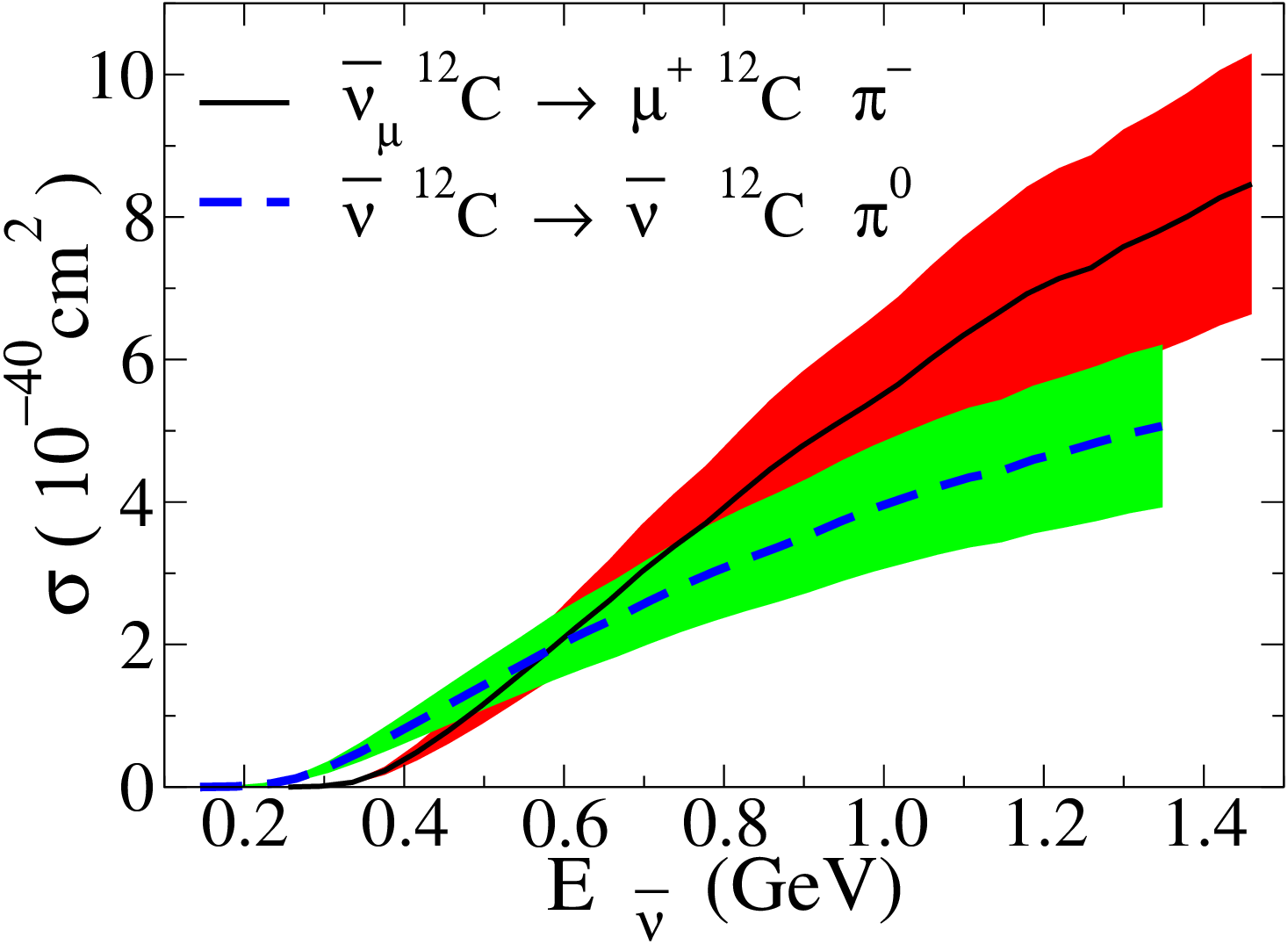}
\hspace{.1cm}\includegraphics[scale=0.25]{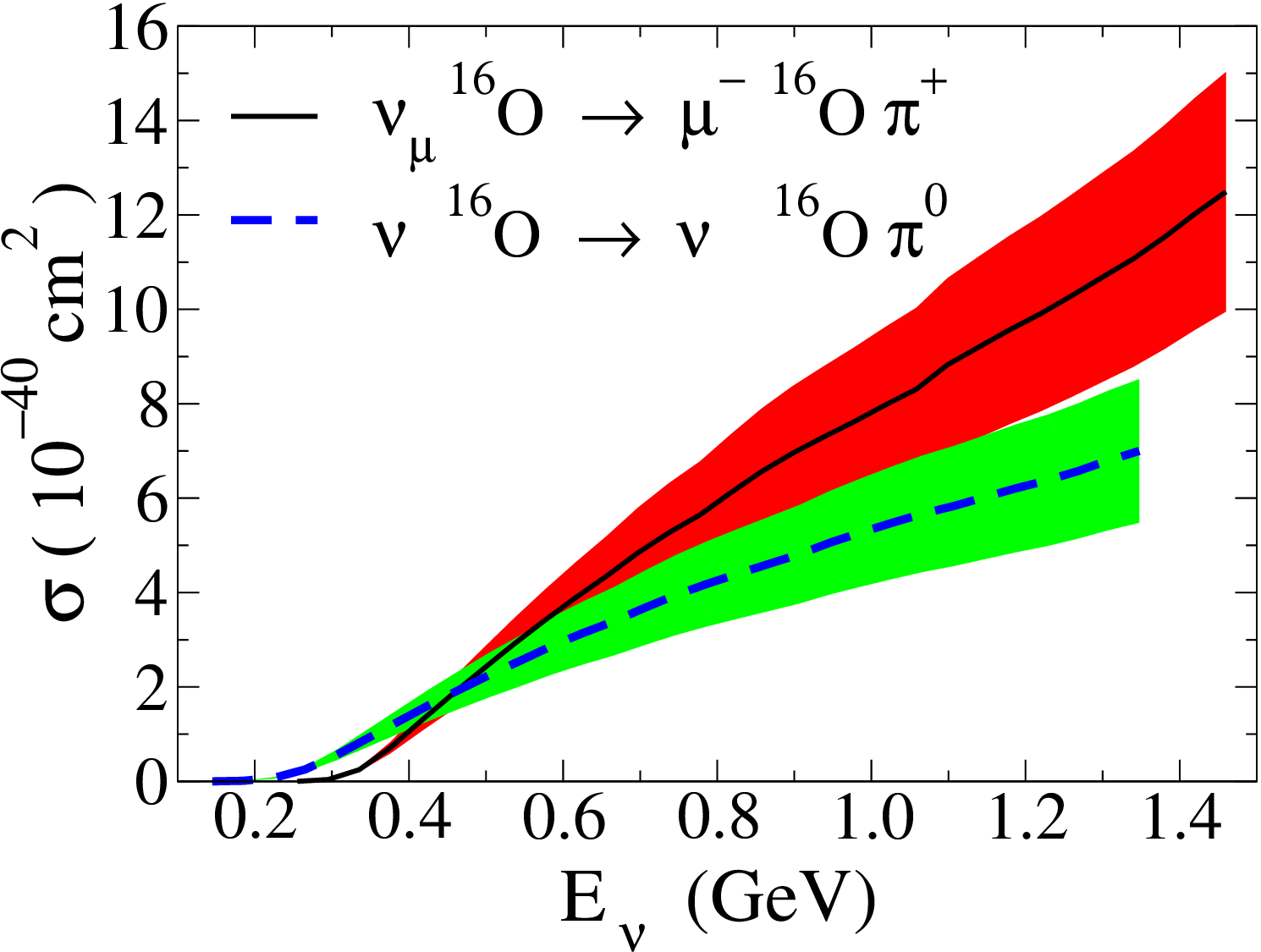}}\\ 
%	  \makebox[0pt]{    \includegraphics[scale=0.3]{secefoxigeno.eps}}
\end{center}
\caption{\footnotesize $\nu_\mu/\bar\nu_\mu$ CC and $\nu/\bar\nu$  NC coherent
pion production from a carbon target (left/middle panel) and $\nu_\mu$
CC and $\nu$ NC coherent pion production from an oxygen target (right panel)
as a function of the neutrino/antineutrino energy.  Error bands are
shown. }\label{fig:cross}
\end{figure}

In Fig.~\ref{fig:cross} we show new $\nu_\mu/\bar\nu$ CC and NC coherent pion 
production total cross sections
off carbon
and oxygen targets. As in Ref.~\cite{Amaro:2008hd}, we observe sizable 
corrections to the approximate relation $\sigma_{\rm CC} \approx 2 
\sigma_{\rm NC}$ for
these two isoscalar nuclei in the whole range of $\nu/\bar\nu$
energies examined. As pointed out in
Refs.~\cite{rein2,Berger:2007rq}, this is greatly due to the
finite muon mass, and thus the deviations are dramatic at low neutrino
energies. In any case, these corrections can not account for the 
apparent incompatibility among  the CC K2K cross section and the NC
value quoted in Ref.~\cite{Raaf}. 

The SciBooNE Collaboration has just reported a measurement of NC
$\pi^0$ production on carbon by a $\nu_\mu$ beam with average
energy 0.8\,GeV~\cite{sciboone}. Based on previous measurements of CC
coherent $\pi^+$ production~\cite{sciboone2}, they conclude that
%
%\begin{equation}
$\left.\frac{\sigma({\rm CC coh}\pi^+)}{\sigma({\rm NC coh}\pi^0)}\right|_{\rm
SciBooNE}=0.14^{+0.30}_{-0.28}$.
%\end{equation}
%
This result can not be accommodated within our model, or any other
present theoretical model, neither microscopic~\cite{luis1,luis2,Martini:2009uj,Nakamura:2009iq}
 nor PCAC based~\cite{rein,rein2,Paschos:2009ag,Berger:2008xs}. Theoretically, 
 this
ratio cannot be much smaller than 1.4-1.6. For instance, for a carbon target and for
a neutrino energy of 0.8\,GeV we find a value of $1.45\pm 0.03$ for that ratio,
ten times bigger that the value given by the SciBooNE Collaboration.
 From the $\nu_\mu+^{12}C$ CC and NC MiniBooNE convoluted results shown
in Table~\ref{tab:res}, we obtain $1.46 \pm 0.03$.  We believe this huge
discrepancy with the SciBooNE result stems form the use in Ref.~\cite{sciboone} of the RS model
to estimate the ratio between NC coherent $\pi^0$ production and the
total CC pion production. As clearly shown in
Refs.~\cite{Amaro:2008hd,Hernandez:2009vm}, the RS model is not
appropriate to describe coherent pion production in the low energy
regime of interest for the SciBooNE experiment.
\vspace*{.15cm}%

\begin{acknowledgments}
  Work supported by DGI and FEDER funds, contracts
  FIS2008-01143/FIS, FIS2006-03438, FPA2007-65748, CSD2007-00042, by JCyL,  
  contracts SA016A07 and GR12,  by GV, contract PROMETEO/2009-0090 and by the
   EU   HadronPhysics2 project, contract 227431. 
\end{acknowledgments}

%\appendix

\end{document}